\documentclass[rnote]{aa} % for a research note

\usepackage{graphicx}
\usepackage{txfonts}

\begin{document}
\title{Star-burst regions in the LMC}

\author{E. Livanou\inst{1}
\and M. Kontizas\inst{1}
\and I. Gonidakis\inst{1}
\and E. Kontizas\inst{2}
\and F. Maragoudaki\inst{1}
\and S. Oliver\inst{3}
\and A. Efstathiou\inst{4}
\and U. Klein\inst{5}}

\offprints{E. Livanou\\
\email{elivanou@phys.uoa.gr}}

\institute{Department of Astrophysics Astronomy \& Mechanics, Faculty
of Physics, University of Athens, GR-15783 Athens, Greece
        \and Institute for Astronomy and Astrophysics, National
Observatory of Athens, P.O. Box 20048, GR-11810 Athens, Greece
         \and Astronomy Centre, Department of Physics \& Astronomy, University of
 Sussex, Brighton, BN1 9QJ, UK
        \and Department of Computer Science and Engineering, Cyprus College, 6 
Diogenes Str, 1516 Nicosia, Cyprus
        \and Radioastronomisches Institut der Universitat, Bonn, Auf dem Hugel 
71, 53121, Bonn, Germany}
        
\date{Received date / accepted}

\abstract
{Aims. Filamentary structures of early type stars are found to be a common feature of 
the Magellanic Clouds formed at an age of about $0.9-2\times10^8\,yr$. As we 
go to younger ages these large structures appear fragmented and sooner or 
later form young clusters and associations. In the optical domain we have 
detected 56 such large structures  of young objects, known as stellar
complexes in the LMC for which we give coordinates and dimensions. We also 
investigate star formation activity and evolution of these stellar complexes 
and define the term ``starburst region''.
Methods. IR  properties of these regions have been investigated using IRAS data. A 
colour-magnitude diagram (CMD) and a two-colour diagram from IRAS data of these 
regions ware compared with observations of starburst galaxies and 
cross-matching with HII regions and SNRs was made . Radio emission maps at 
8.6-GHz and the CO (1$\rightarrow$0) line were also cross correlated with the 
map of the stellar complexes.
Results. It has been found that nearly 1/3 of the stellar complexes are extremely 
active resembling the IR behaviour of starburst galaxies and HII regions. 
These stellar complexes illustrating such properties are called here 
``starburst regions''. They host an increased number of HII regions and SNRs. 
The main starburst tracers are their IR luminosity ($F_{60}$ well above 5.4 
Jy) and the 8.6-GHz radio emission. Finally the evolution of all stellar 
complexes is discussed based on the CO emission.
}
\maketitle

\section{Introduction}

Due to its proximity to our Galaxy and  the SMC, the LMC offers two important 
advantages: i) the possibility to study its stellar content down to low 
stellar masses and ii) the possibility to follow the consequences of the close 
interaction, about $2-4 \times 10^8\,yr$ ago (Gardiner \& Noguchi 
\cite{gardiner}; Kunkel, Demers \& Irwin \cite{kunkel}), with the SMC. Stellar complexes 
are defined as large  structures (clump like) in galaxies dominated by recently
 formed stellar component mixed with very young clusters, stellar associations 
and gas (Martin et al. \cite{martin}; van den Bergh \cite{van}; Elmegreen  
\& Elmegreen \cite{elmegreen2}; Feitzinger \& Braunsfurth \cite{feitz}; Elmegreen 
\& Elmegreen \cite{elmegreen3}; Ivanov \cite{ivanov}; Larson et al. \cite{larson1}; 
Efremov\cite{efremov1}; Larson et al. \cite{larson2}; Efremov \& Chernin 
\cite{efremov3}; Elmegreen et al. \cite{elmegreen4}; Kontizas et al. 
\cite{kontizas1}; Efremov \cite{efremov2}; Battinelli, Efremov \& Magnier 
\cite{battinelli}; Elmegreen \& Efremov \cite{elmegreen1}; Kontizas et al. 
\cite{kontizas2}). So far they have been detected as regions with young 
stellar component and enhanced stellar number density, compared to the 
surrounding region. In the LMC  stellar complexes have been detected using UK 
Schmidt plates (Maragoudaki et al. \cite{maragoudaki1}).
In the optical domain we have detected 56  
large  structures  of young objects. These stellar complexes are associated 
with the already known nine Shapley constellations (Shapley \cite{shapley}), 
which are referred to as large stellar structures of early type  supergiants. 

Helou (\cite{helou}) found that the IRAS 
$log(F_{60}/F_{100})$ versus $log(F_{12}/F_{25})$ colour-colour diagram of 
``normal galaxies'' shows a distribution that extends continuously from the 
cool ($\sim$20K) relatively constant ``cirrus'' emission from the neutral
medium to the warmer ($\sim$40-50K) emission from the active, starburst end.

IRAS far-IR fluxes for 6 HII regions in the LMC have been studied by
DeGioia-Eastwood (\cite{gioia}). These regions are the sites of massive star 
formation, where the radiative heating source is young stars rather than the 
general interstellar radiation field. Such regions are expected to lie in 
stellar complexes and/or to be the stage just before the stellar complex is 
formed. 

The highly structured diffuse X-ray emission of the LMC has been imaged in 
detail by ROSAT. The brightest regions are found east of LMC X-1  and in the 
30 Doradus region. The latter is strikingly similar to the optical picture. 
There is a strong correlation between diffuse features in the X-ray image and 
ESO colour images of the LMC in visible light. Bright knots in the X-ray map 
correspond to HII emission in the optical (Westerlund \cite{west}). The 
distribution of the SNRs in the LMC shows (i) a clumping of objects in the 30 
Doradus region, (ii) several remnants within the Bar of the LMC and (iii) that 
the remainder of the remnants are found in super-associations. Most SNRs in the 
Bar are in regions where many young clusters are located, clusters as young as 
$\le\,10\,Myr$ (Westerlund \cite{west}).

The LMC has been well studied in the radio continuum, which is connected to
the formation of new and massive stars. Fukui et al. (\cite{fukui1}, \cite{fukui2}) 
found that the molecular clouds in the LMC have a good correlation with the youngest 
($\la$10~Myr) stellar clusters, while there is little spatial correlation of 
these clouds with SNRs or with the older stellar clusters. 

In this paper we describe how the stellar complexes were detected, using 
optical data and we give their location and dimensions. The total flux of each 
structure is presented. 
Their derived IR  properties are compared to IR 
properties of galaxies in order to identify the distinction between ordinary 
stellar complexes and active star formation regions. Possible connection of 
the complexes determined here with known X-ray detected SNRs and candidates as 
well as HII regions is discussed. Finally, their correlation between high 
resolution radio data at 8.6-GHz and CO emission is examined, in order to 
determine the activity of the complexes and their evolution.

\section{Optical data}

The large-scale structures in nearby galaxies require extended field 
observational material in order to be treated homogeneously. Therefore 
we used direct photographic plates taken 
with the UK 1.2m Schmidt Telescope, in various wavebands: U, R and HeII 
($\lambda=4686$ \AA) down to a magnitude limit of $19-20\,mag$. 
These plates were digitised using the fast measuring
machines APM  and SuperCosmos. The derived data were produced as catalogues of
detected stellar images. Star counts were performed on the above catalogues 
in order to derive isopleth contour maps of the LMC in the various wavebands 
and magnitude slices, to trace the differences between faint and brighter 
stellar populations. 
Star counts, in the optical wavelength range, of large areas in the LMC have 
revealed 56 groupings of young stars, known as stellar complexes (with 
dimensions from 150\,$pc$ to $\sim$\,1500\,$pc$). Their coordinates, 
dimensions and properties are given in Table \ref{Properties}, whereas their 
ages span from a few $\times 10^8\,yr$ to very young ($\sim 10^6\,yr$). The 
ages are derived from their stellar component, as described by Maragoudaki et
al. (\cite{maragoudaki1}), which consist of early type stars (O, B, A), 
dust, young clusters and associations, very often in groups. The parental 
density fluctuations, which were fragmented into stellar complexes can be 
explained as a result of the close encounter of the SMC with the LMC that 
happened a few $\times 10^8\,yr$ ago (Gardiner \& Noguchi \cite{gardiner}; 
Kunkel et al. \cite{kunkel}).

\section{IR data}

IRAS all-sky survey Images taken at 12, 25, 60 and 100\,$\mu$m have been
obtained from SkyView (http://skyview.gsfc.nasa.gov/skyview.html) for the 
whole area of the LMC to study the IR properties of the stellar complexes 
detected in the optical domain.
From the IRAS catalogues we calculated a total flux for each complex as the 
sum of the fluxes corresponding to the pixels, which cover the predetermined 
complex area. These total fluxes were corrected from the background.
The
background value was assumed as the average of four flux/pixel values at the 
four rather clear corners of each frame, for all IRAS wavelengths
respectively. The background corrected total 
fluxes were averaged to a mean value (flux/pixel in MJy/sr) for all 
complexes so that they can be more easily compared with the literature.

\begin{figure}
\centering
\includegraphics[angle=-90,width=8cm]{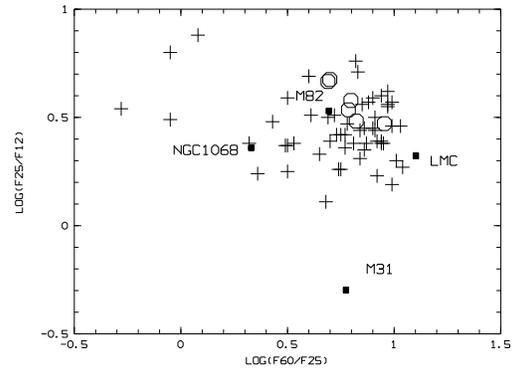}
\caption{IRAS colour-colour diagram for the detected stellar complexes. The 
black-filled squares mark the observed colours of the LMC (irregular galaxy)
and three typical galaxies: M31 (disk), M82 (starburst) and NGC1068 (AGN). The 
circles mark LMC HII regions.}
\label{c-c-1}
\end{figure}

Fig.~\ref{c-c-1} shows  
the IRAS colour-colour diagram of 
$log(F_{25}/F_{12})$ vs $log(F_{60}/F_{25})$  
for the detected stellar complexes. The black-filled 
squares represent the integrated values for NGC1068, M82, M31 and LMC 
respectively, indicating the loci of different types of galaxies on this 
diagram. NGC1068 is a prototype AGN (Rowan-Robinson \& Crawford \cite{rowan1}),
M82 is a prototype starburst galaxy, while M31 is a typical disk galaxy (Rice, 
W., et al. \cite{rice}). The LMC HII regions of DeGioia-Eastwood 
(\cite{gioia}) are plotted with circles.
We notice that many stellar complexes lie around 
the location of M82, a prototype starburst, and the HII regions revealing 
enhanced star formation activity.

\begin{figure}
\centering
\includegraphics[angle=-90,width=8cm]{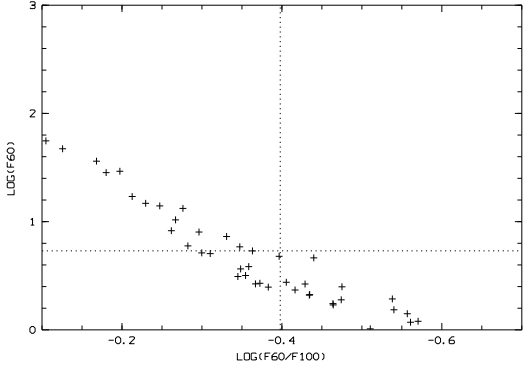}
\caption{IRAS flux density in $logF_{60}$ versus IRAS colour 
$log(F_{60}/F_{100})$ for all the structures detected here. Two lines are 
plotted which represent $F_{60}=5.4\, Jy$ and $F_{60}/F_{100}=0.4$. The 
upper left area indicates the loci of very active star forming regions.}
\label{Lum60}
\end{figure}

The IRAS flux densities in $logF_{60}$ versus the IRAS colour 
$log(F_{60}/F_{100})$ for all the large structures detected here are 
illustrated in Fig.~\ref{Lum60}. Lehnert \& Heckman (\cite{lehnert}) seeking to improve our 
understanding of the range of physical processes occurring within IR selected 
galaxies, found that the starburst galaxies are IR ``warm'' if 
$F_{60}/F_{100}\ge 0.4$ ($log(F_{60}/F_{100})\ge -0.39$) and IR ``bright'' 
if $F_{60}\ge 5.4$ Jy ($log(F_{60})\ge 0.73$). From  Fig.~\ref{Lum60} it is 
obvious that nearly 1/3 of the detected stellar complexes fulfil the criteria 
defined by Lehnert \& Heckman (\cite{lehnert}) for the starburst galaxies. 
The three upper points correspond to regions in 30 Doradus (complexes C17, 
A21, A24), where there is the highest concentration of SNRs. The 15 structures 
that have $log(F_{60}/F_{100})\ge -0.39$ and $log(F_{60})\ge 0.73$, resemble 
the behaviour of starburst galaxies, whereas the rest 41 do not show enhanced 
star formation activity.

\section{Radio and CO data}

\begin{figure}
\centering
\includegraphics[angle=0,width=8cm]{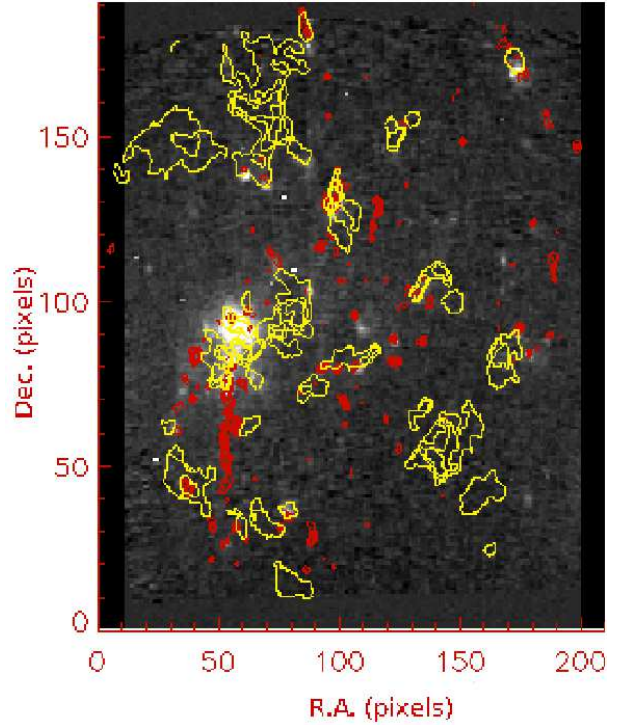}
\caption{Maps of the LMC at 8.6~GHz (grey-scale) and in the
CO(1$\rightarrow$0) line (red contours), with the stellar complexes 
superimposed with yellows lines.}
\label{co8.6}
\end{figure}

We used the recently published 8.6-GHz map of Dickel et al. (\cite{dickel}),
with its angular resolution of 20\arcsec\ and linear resolution $\sim$5~pc 
and the CO (1$\rightarrow$0) line that was fully mapped by Fukui et al. 
(\cite{fukui1}), with angular resolution of 2.6\arcmin\ corresponding to a 
spatial resolution of $\sim$40~pc. In Fig.~\ref{co8.6} we have superimposed 
the stellar aggregates, complexes and super-complexes (yellow lines) onto the
radio 8.6-GHz (grey-scale) and CO (red contours) map. The time sequence evident 
in this figure is such that the molecular gas (CO) traces regions of ongoing and 
(near-) future star formation, the ionised gas (8.6-GHz) locates ongoing and 
somewhat evolved star formation ($10^7$ $\sim$ $10^8$~yr). We examined the 
correlation between 8.6-GHz radio emission, the CO emission and the location 
of the 56 stellar complexes (Table ~\ref{Properties}, columns 8 and 9).

We notice that the majority (13 out of 15) of the regions 
with $F_{60}>5.4 Jy$, coincide well with the 8.6-GHz radio emission. The 
existence of CO emission in seven of them, reveals future star formation 
activity. The rest 2, that lack both 8.6-GHz and CO radiation, are actually 
found to have values of $F_{60}$ very close to $5.4 Jy$.
Considering the 41 regions with $F_{60}<5.4 Jy$, the opposite behaviour is 
revealed. 36 out of 41 regions do not exhibit significant amount of 8.6-GHz 
radiation, while the remaining 5 that do exhibit 8.6-GHz radiation, are found 
to have $F_{60}$ very close to $5.4 Jy$.   

The CO emission is indicative of on-going and possible future star formation 
activity and there is no pattern revealed regarding the $F_{60}$ of the 
complexes. 16 out of the 56 complexes have significant CO emission, 
indicating possible future star formation activity and only 10 of them are
correlated with 8.6-GHz emission as well.

\section{Discussion \& Conclusions}

From optical observations, 56 stellar complexes are revealed in an area of 
$6\times 7 \deg^{2}$ of the LMC. The IR properties are studied using the IRAS 
fluxes. The ratio of the flux densities $F_{60}/F_{100}$, the flux density 
$F_{60}$ as discussed above, the 8.6-GHz and CO radio emission, will be used
in order to classify these complexes. 

As mentioned in section 3, Lehnert \& Heckman (\cite{lehnert}) showed that
starburst galaxies are IR ``warm'' and ``bright''. In the same manner, we 
adopt the equivalent term {\bf ``starburst regions''} for complexes that are 
found to fulfil these criteria, namely $F_{60}/F_{100} \ge 0.4$ and $F_{60} 
\ge 5.4$ Jy. In addition, complexes that do not fulfil the previously stated
criteria will be called {\bf ``active complexes''}.    

A more detailed study of high resolution 8.6-GHz map (Fig.~\ref{co8.6}) 
revealed that all complexes with F$_{60}$ well above the 5.4 Jy limit are 
very well correlated with the 8.6-GHz radio emission. In contrast, complexes 
with F$_{60}$ well below the 5.4 Jy limit lack 8.6-GHz emission. 
However, 7 out of the 56 complexes (A5, A6, A7, C5, C7, C22 and A31) have 
F$_{60}$ close to the classification limit and deviate from the above 
behaviour. Complex C22, for example, is well correlated with 8.6-GHz emission;
its {\it F$_{60}$}(5.07$\pm0.20$ Jy) though is below the accepted limit. 
In contrast A31 has {\it F$_{60}$=5.98$\pm0.20$ Jy}, which would classify 
it as a starburst region; however, there is no 8.6-GHz emission associated 
with it. Since the $F_{60}$ criterion is not trustworthy for these regions, 
we additionally use the 8.6-GHz emission to classify the complexes as 
``starburst candidates'' or ``active complexes candidates'' depending on its 
existence. Hence, using the 8.6-GHz radiation as an additional tracer in 
tandem with the $F_{60}$, we can classify the stellar complexes into 13 
starbursts, 5 starburst candidates, 2 active complexes candidates and 36 
active complexes as seen in Table ~\ref{starbursts}. The characterisation of
each complex is given in Table ~\ref{Properties}, column 10.

\begin{table}
\begin{center}
\caption{Complexes characterisation.}
\begin{tabular}{lccc}
\hline \hline
Complex Type             & $F_{60}$        & 8.6      &  No. of  \\
                         & (Jy)            & (GHz)    & Complexes \\
\hline
starburst                & $F_{60}>$5.4    & yes      &   13              \\
starburst candidate      & $F_{60}\la$5.4  & yes      &   5               \\
active complex candidate & $F_{60}\ga$5.4  & no       &   2               \\
active complex           & $F_{60}<$5.4    & no       &   36              \\ 
\hline
\end{tabular}
\label{starbursts}
\end{center}
\end{table}

Moreover, a discrimination concerning the evolution of all the complexes was
attempted based on the CO data. More than half of the starburst and starburst
candidate regions show enhanced CO emission indicating ongoing and future 
evolution, while the rest of them are thought to be more or less evolved. 
Regarding the active stellar complexes (and the two candidates), 
we found only 5 of them with significant amount of CO emission, giving 
additional indication of current star formation and revealing potential future
starbursts. The lack of CO in the rest of them could indicate that these
complexes have never been starbursts or that their starburst activity 
previously exhausted the molecular gas.

\begin{figure}
\centering
\includegraphics[angle=0,width=8cm]{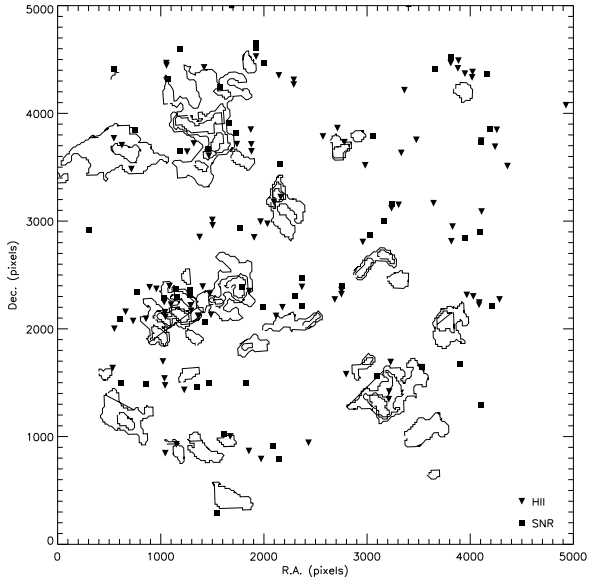}
\caption{The HII regions (triangles) and SNRs (squares), plotted over the 
stellar complexes as defined here.}
\label{SBR}
\end{figure}  

In Fig.~\ref{SBR} we plotted the known SNRs and HII regions based on the 
literature (Williams et al. \cite{williams}; Haberl \& Pietsch \cite{haberl}; 
Filipovic et al. \cite{filip} and Sasaki, Haberl \& Pietsch \cite{sasaki}) over the 
stellar complexes with triangles and squares respectively.
The overplotted HII regions show an increased concentration 
in and around the very active star forming regions, as expected. 
Most of the defined starburst areas (and the candidates) are associated with 
SNRs. The detected complexes are 
also cross-matched with the catalogue of stellar associations (Gouliermis et 
al. \cite{gouliermis}) and that of nebulae in the LMC (Davies, Elliot \& Meaburn 
\cite{davies}). The majority of the complexes are loci of stellar 
associations, whereas all of them are associated with a large number of 
nebulosities (Table ~\ref{Properties}).

\begin{table*}[h]
\begin{center}
\tiny{
\caption{List of the LMC stellar complexes. Col. 1 indicates the adopted 
identification name. A denotes aggregate, C complex and SC super-complex. For 
a structure found inside a larger one, the name of the later is given in 
parenthesis. Col. 2, 3 and 4 give the RA, DEC of the ``centre'' of each 
stellar complex and its dimension. Col. 5,6, 7 show the number of SNRs and 
candidates, stellar associations and number of Nebulae and Henize objects 
found in each complex. Col. 8 and 9 indicate presence of 8.6-GHz and 
CO line emission respectively. Finally in Col. 10 the complexes are
characterised. I indicates starbursts, II starburst candidates, III active
complexes candidates and IV active complexes.}
\begin{tabular}{lccccccccc}
Stellar grouping & RA (1950) & DEC (1950) & Dimension & No.SNRs &Association
                  ID & No.Nebulae & 8.6 & CO & Characterisation\\
                  & h   m   s  & deg  m  s  & (pc)&  & & & GHz &line & \\ 
\hline \hline
SC1       & 05:44:39 & -67:15:18 & 1421.5 & 1            & 173,257     & 46 & -       & -       & IV \\
C1 (SC1)  & 05:43:30 & -67:17:29 & 355.3  & 0            & 173,257     & 57 & -       & -       & IV \\
A1        & 05:52:53 & -67:29:46 & 205    & 0            & 173,257     & 22 & -       & -       & IV \\
A2        & 05:43:44 & -67:19:37 & 191    & 1            & 173,257     & 31 & -       & -       & IV \\
SC2       & 05:31:22 & -66:54:42 & 1526.5 & 7            &             & 24 & -       & -       & IV \\
C2 (SC2)  & 05:32:45 & -67:05:11 & 749    & 1            & 173         & 41 & -       & -       & IV \\
C3 (C2)   & 05:32:47 & -66:58:22 & 572    & 1            &             & 32 & -       & -       & IV \\
A3 (SC2)  & 05:29:45 & -66:37:17 & 204    & 0            &             & 31 & -       & -       & IV \\
A4 (C2)   & 05:30:53 & -66:39:15 & 245    & 0            &             & 26 & -       & -       & IV \\
A5 (SC2)  & 05:32:55 & -67:32:29 & 164    & 1            & 173,257     & 41 & -       & -       & III\\
A6        & 05:25:58 & -66:12:18 & 299    & 1            &             & 20 & $\surd$ & $\surd$ & II \\
A7        & 05:27:01 & -67:27:42 & 299    & 1            & 173,257     & 10 & $\surd$ & -       & II \\
C4        & 05:13:28 & -67:20:57 & 314    & 0            & 173,257     & 16 & -       & -       & IV \\
A8 (C4)   & 05:13:10 & -67:22:34 & 246    & 0            & 173,257     & 16 & -       & -       & IV \\
A9        & 05:10:55 & -67:11:35 & 192    & 0            & 173,257     & 26 & -       & $\surd$ & IV \\
A10       & 04:57:01 & -66:28:27 & 245    & 1            &             & 7  & $\surd$ & $\surd$ & I  \\
C5        & 05:20:58 & -68:10:16 & 765    & 2            &             & 17 & $\surd$ & $\surd$ & II \\
C6 (C5)   & 05:22:13 & -67:58:26 & 342.6  & 1            & 173,257     & 36 & $\surd$ & $\surd$ & I  \\
A11 (C6)  & 05:22:13 & -67:56:44 & 205    & 1            & 173,257     & 17 & $\surd$ & $\surd$ & I  \\
C7        & 05:07:30 & -68:44:27 & 545    & 1            &             & 39 & $\surd$ & $\surd$ & II \\
A12 (C7)  & 05:09:04 & -68:49:37 & 287    & 1            &             & 17 & $\surd$ & $\surd$ & I  \\
A13 (C7)  & 05:05:38 & -68:39:12 & 245    & 0            &             & 27 & -       & -       & IV \\
A14       & 05:04:00 & -68:57:45 & 230    & 0            &             & 11 & -       & -       & IV \\
C8        & 04:55:33 & -69:32:52 & 628    & 1            &             & 12 & -       & -       & IV \\
A15 (C8)  & 04:56:55 & -69:30:18 & 273    & 1            &             & 10 & -       & -       & IV \\
C9        & 05:09:26 & -70:08:58 & 328    & 0            & 150,153,151 & 19 & -       & -       & IV \\
A16 (C9)  & 05:08:09 & -70:05:05 & 246    & 0            & 150,151     & 10 & -       & -       & IV \\
SC3       & 05:04:44 & -70:29:23 & 1093.2 & 2            & 150,153,151 & 23 & -       & -       & IV \\
C10 (SC3) & 05:05:01 & -70:32:56 & 478.3  & 1            & 150,153,151 & 13 & -       & -       & IV \\
C11(C10)  & 05:06:16 & -70:40:18 & 328    & 0            & 150,153,151 & 17 & -       & -       & IV \\
C12       & 04:58:02 & -70:53:27 & 574    & 0            &             & 2  & -       & -       & IV \\
A17       & 04:56:28 & -71:28:37 & 164    & 0            &             & 0  & -       & -       & IV \\
C13       & 05:30:55 & -71:50:37 & 574    & 0            & 76          & 6  & -       & -       & IV \\
C14       & 05:39:55 & -71:11:22 & 328    & 0            & 76          & 10 & -       & $\surd$ & IV \\
C15       & 05:35:42 & -71:12:22 & 437    & 0            & 76          & 17 & -       & $\surd$ & IV \\
A18       & 05:31:26 & -71:06:11 & 218.64 & 2            & 76          & 21 & $\surd$ & $\surd$ & I  \\
C16       & 05:47:50 & -70:41:07 & 615    & 1 very close & 150,151     & 15 & -       & $\surd$ & IV \\
A19 (C16) & 05:47:53 & -70:44:31 & 210    & 0            & 150,153,151 & 14 & -       & $\surd$ & IV \\
A20       & 05:49:53 & -70:05:51 & 164    & 0            & 150,153,151 & 10 & -       & -       & IV \\
C17       & 05:39:42 & -69:21:58 & 724    & 10           & 172         & 23 & $\surd$ & $\surd$ & I  \\
C18 (C17) & 05:38:17 & -69:26:16 & 463    & 3            & 172         & 50 & $\surd$ & -       & I  \\
A21 (C17) & 05:36:42 & -69:10:51 & 272.3  & 5            & 172         & 29 & $\surd$ & -       & I  \\
A22 (C18) & 05:39:34 & -69:28:00 & 218    & 1            & 172         & 47 & $\surd$ & -       & I  \\
A23 (C18) & 05:36:59 & -69:24:30 & 245    & 1            & 172         & 32 & $\surd$ & -       & I  \\
A24 (C17) & 05:40:52 & -69:38:12 & 298    & 2            & 172         & 44 & $\surd$ & -       & I  \\
A25       & 05:35:48 & -68:55:28 & 190.61 & 0            &             & 29 & $\surd$ & $\surd$ & I  \\
A26       & 05:35:18 & -69:43:05 & 231.45 & 0            & 172         & 46 & $\surd$ & -       & I  \\
A27       & 05:36:33 & -70:10:26 & 272.3  & 0            & 150,153,151 & 32 & -       & -       & IV \\
C19       & 05:30:00 & -69:11:52 & 792.57 & 2            & 172         & 10 & -       & -       & IV \\
C20 (C19) & 05:29:42 & -69:08:07 & 517.37 & 2            & 172         & 46 & -       & -       & IV \\
A28 (C20) & 05:31:34 & -69:15:19 & 163    & 0            & 172         & 63 & -       & -       & IV \\
A29 (C20) & 05:28:27 & -69:04:25 & 285    & 2            & 172         & 71 & -       & -       & IV \\
C21       & 05:26:03 & -69:51:28 & 463    & 1 very close & 172         & 36 & -       & -       & IV \\
A30 (C21) & 05:27:03 & -69:50:03 & 163    & 0            & 172         & 28 & -       & -       & IV \\
C22       & 05:19:27 & -69:34:39 & 763    & 2            & 177         & 38 & $\surd$ & $\surd$ & II \\
A31 (C22) & 05:17:09 & -69:31:39 & 300    & 0            & 772         & 20 & -       & -       & III\\ 
\hline
\end{tabular}
\label{Properties}
}
\end{center}
\end{table*}

\begin{table*}[h]
\begin{center}
\tiny{
\caption{List of the derived IRAS fluxes (MJy/sr per pixel), luminocities 
(in units of solar luminocities) of the identified stellar groupings at 12,
 25, 60 and 100 $\mu m$, the IRAS ratio $F_{60}/F_{100}$ and the flux at 60\,
$\mu m$ in Jy per pixel.}
\begin{tabular}{lcccccccccc}
Stellar grouping & f12 & L12 & f25 & L25 & f60 & L60 & f100 & L100 & $F_{60}/F_{100}$ & $ F_{60} $   \\
\hline \hline
SC1          & 0.18 & 2.8E+06 &  0.50 & 3.7E+06 &   3.63 & 1.1E+07 &   8.87 & 1.7E+07 & 0.41 &  0.69 $ \pm $ 0.25 \\
C1 (SC1)     & 0.31 & 0.3E+06 &  1.02 & 0.5E+06 &   5.38 & 1.1E+06 &  13.38 & 1.6E+06 & 0.40 &  1.02 $ \pm $ 0.23 \\
A1           & 0.36 & 0.1E+06 &  1.28 & 0.2E+06 &   0.67 & 0.4E+06 &   2.16 & 0.8E+06 & 0.31 &  0.13 $ \pm $ 0.49 \\
A2           & 0.47 & 0.1E+06 &  1.54 & 0.2E+06 &   6.32 & 0.4E+06 &  19.39 & 0.7E+06 & 0.33 &  1.20 $ \pm $ 0.23 \\
SC2          & 1.14 & 2.0E+07 &  4.46 & 3.8E+07 &  13.96 & 5.0E+07 &  33.50 & 7.2E+07 & 0.42 &  2.65 $ \pm $ 0.21 \\
C2 (SC2)     & 0.16 & 0.7E+06 &  0.56 & 1.1E+06 &   4.03 & 3.5E+06 &   9.34 & 4.9E+06 & 0.43 &  0.77 $ \pm $ 0.24 \\
C3 (C2)      & 0.11 & 0.3E+06 &  0.52 & 0.6E+06 &   2.08 & 1.1E+06 &   5.13 & 1.5E+06 & 0.41 &  0.40 $ \pm $ 0.29 \\
A3 (SC2)     & 0.17 & 0.5E+06 &  1.10 & 0.2E+06 &   0.98 & 0.6E+05 &   3.51 & 0.1E+06 & 0.28 &  0.19 $ \pm $ 0.40 \\
A4 (C2)      & 0.12 & 0.6E+05 &  0.89 & 0.2E+06 &   1.06 & 0.1E+06 &   3.46 & 0.2E+06 & 0.31 &  0.20 $ \pm $ 0.38 \\
A5 (SC2)     & 1.39 & 0.3E+06 &  3.80 & 0.4E+06 &  30.79 & 1.3E+06 &  64.60 & 0.6E+06 & 0.48 &  5.85 $ \pm $ 0.20 \\
A6           & 1.16 & 0.8E+06 &  2.80 & 0.9E+06 &  25.17 & 3.4E+06 &  58.76 & 4.8E+06 & 0.43 &  4.78 $ \pm $ 0.20 \\
A7           & 0.82 & 0.6E+06 &  2.89 & 0.9E+06 &  27.08 & 3.7E+06 &  50.10 & 4.1E+06 & 0.54 &  5.15 $ \pm $ 0.20 \\
C4           & 0.68 & 0.5E+06 &  1.53 & 0.6E+06 &  11.15 & 1.7E+06 &  26.38 & 0.4E+06 & 0.42 &  2.12 $ \pm $ 0.21 \\
A8 (C4)      & 0.80 & 0.4E+06 &  1.90 & 0.4E+06 &  12.30 & 1.1E+06 &  28.14 & 1.5E+06 & 0.44 &  2.34 $ \pm $ 0.21 \\
A9           & 0.52 & 0.1E+06 &  1.66 & 0.2E+06 &   8.07 & 0.5E+06 &  23.91 & 0.8E+06 & 0.34 &  1.53 $ \pm $ 0.22 \\
A10          & 2.26 & 1.0E+06 &  8.83 & 1.9E+06 &  69.77 & 6.5E+06 & 127.91 & 7.1E+06 & 0.55 & 13.26 $ \pm $ 0.20 \\
C5           & 1.61 & 7.2E+06 &  4.71 & 1.0E+07 &  28.23 & 2.5E+07 &  61.20 & 3.3E+07 & 0.46 &  5.36 $ \pm $ 0.20 \\
C6 (C5)      & 0.10 & 0.9E+05 &  6.66 & 2.9E+06 &  54.65 & 9.9E+06 &  97.19 & 1.1E+07 & 0.56 & 10.38 $ \pm $ 0.20 \\
A11 (C6)     & 3.21 & 1.0E+06 & 11.89 & 1.8E+06 &  89.85 & 5.8E+06 & 142.78 & 5.6E+06 & 0.63 & 17.07 $ \pm $ 0.20 \\
C7           & 0.96 & 2.2E+06 &  2.31 & 2.5E+06 &  20.22 & 9.3E+06 &  42.25 & 1.2E+07 & 0.49 &  3.84 $ \pm $ 0.21 \\
A12 (C7)     & 1.89 & 1.2E+06 &  5.31 & 1.6E+06 &  42.20 & 5.4E+06 &  79.59 & 6.1E+06 & 0.53 &  8.02 $ \pm $ 0.20 \\
A13 (C7)     & 0.59 & 0.3E+06 &  1.56 & 0.4E+06 &   9.20 & 0.8E+06 &  22.76 & 1.3E+06 & 0.40 &  1.75 $ \pm $ 0.22 \\
A14          & 0.94 & 0.4E+06 &  1.90 & 0.4E+06 &  13.16 & 1.1E+06 &  35.27 & 1.7E+06 & 0.37 &  2.50 $ \pm $ 0.21 \\
C8           & 0.90 & 2.7E+06 &  2.44 & 3.6E+06 &  16.74 & 1.0E+07 &  33.96 & 1.2E+07 & 0.49 &  3.18 $ \pm $ 0.21 \\
A15 (C8)     & 0.81 & 0.4E+06 &  1.86 & 0.5E+06 &  11.02 & 1.3E+06 &  25.97 & 1.8E+06 & 0.42 &  2.09 $ \pm $ 0.21 \\
C9           & 1.20 & 1.0E+06 &  3.67 & 1.5E+06 &   9.98 & 1.7E+06 &  25.75 & 2.6E+06 & 0.39 &  1.90 $ \pm $ 0.22 \\
A16 (C9)     & 0.72 & 0.3E+06 &  1.77 & 0.4E+06 &   8.95 & 0.8E+06 &  22.08 & 1.2E+06 & 0.41 &  1.70 $ \pm $ 0.22 \\
SC3          & 0.40 & 3.7E+06 &  0.87 & 3.9E+06 &   3.87 & 7.1E+06 &   9.74 & 1.1E+07 & 0.40 &  0.74 $ \pm $ 0.25 \\
C10 (SC3)    & 0.46 & 0.8E+06 &  1.07 & 0.9E+06 &   3.38 & 1.2E+06 &   8.85 & 1.9E+06 & 0.38 &  0.64 $ \pm $ 0.25 \\
C11 (C10)    & 0.60 & 0.5E+06 &  1.44 & 0.6E+06 &   4.82 & 0.8E+06 &  12.47 & 1.2E+06 & 0.39 &  0.92 $ \pm $ 0.24 \\
C12          & 0.37 & 0.9E+06 &  0.87 & 1.1E+06 &   1.82 & 0.9E+06 &   5.75 & 1.8E+06 & 0.32 &  0.35 $ \pm $ 0.30 \\
A17          & 0.69 & 0.1E+06 &  2.14 & 0.2E+06 &   1.89 & 0.8E+05 &   7.51 & 0.2E+06 & 0.25 &  0.36 $ \pm $ 0.30 \\
C13          & 0.45 & 1.1E+06 &  0.79 & 9.6E+05 &   2.51 & 1.3E+06 &   8.61 & 2.6E+06 & 0.29 &  0.48 $ \pm $ 0.27 \\
C14          & 0.99 & 0.8E+06 &  1.81 & 0.7E+06 &  10.17 & 1.7E+06 &  31.07 & 3.1E+06 & 0.33 &  1.93 $ \pm $ 0.22 \\
C15          & 0.62 & 0.9E+06 &  1.13 & 0.8E+06 &   6.19 & 1.8E+06 &  18.46 & 3.3E+06 & 0.34 &  1.18 $ \pm $ 0.23 \\
A18          & 1.86 & 0.7E+06 &  4.59 & 0.8E+06 &  38.36 & 2.8E+06 &  78.27 & 3.5E+06 & 0.49 &  7.29 $ \pm $ 0.20 \\
C16          & 0.67 & 1.9E+06 &  0.87 & 1.2E+06 &   4.16 & 2.4E+06 &  17.37 & 6.1E+06 & 0.24 &  0.79 $ \pm $ 0.24 \\
A19 (C16)    & 1.08 & 0.4E+06 &  1.86 & 0.3E+06 &   4.24 & 0.3E+06 &  21.73 & 0.9E+06 & 0.20 &  0.81 $ \pm $ 0.24 \\
A20          & 0.05 & 0.1E+05 &  0.01 & 0.1E+04 &  13.07 & 0.5E+06 &  27.62 & 0.7E+06 & 0.47 &  2.48 $ \pm $ 0.21 \\
C17          & 7.21 & 2.9E+07 & 36.90 & 7.2E+07 & 248.00 & 2.0E+08 & 327.45 & 1.6E+08 & 0.76 & 47.12 $ \pm $ 0.20 \\
C18 (C17)    & 3.88 & 6.4E+06 & 16.18 & 1.3E+07 & 149.37 & 4.9E+07 & 222.33 & 4.4E+07 & 0.67 & 28.38 $ \pm $ 0.20 \\
A21 (C7)     & 7.62 & 4.3E+06 & 43.93 & 1.2E+07 & 293.57 & 3.4E+07 & 369.90 & 2.5E+07 & 0.79 & 55.78 $ \pm $ 0.20 \\
A22 (C18)    & 4.26 & 1.6E+06 & 15.66 & 2.8E+06 & 153.80 & 1.1E+07 & 238.40 & 1.0E+07 & 0.65 & 29.22 $ \pm $ 0.20 \\ 
A23 (C18)    & 2.30 & 1.1E+06 &  8.35 & 1.9E+06 &  77.75 & 7.2E+06 & 128.05 & 7.1E+06 & 0.61 & 14.77 $ \pm $ 0.20 \\
A24          & 5.50 & 3.8E+06 & 21.67 & 7.1E+06 & 190.57 & 2.6E+07 & 276.81 & 2.3E+07 & 0.69 & 36.20 $ \pm $ 0.20 \\
A25          & 1.42 & 0.4E+06 &  4.05 & 0.5E+06 &  43.42 & 2.4E+06 &  75.47 & 2.5E+06 & 0.57 &  8.25 $ \pm $ 0.20 \\
A26          & 2.62 & 1.1E+06 &  7.49 & 1.5E+06 &  73.60 & 6.1E+06 & 126.18 & 6.3E+06 & 0.58 & 13.98 $ \pm $ 0.20 \\
A27          & 1.64 & 0.9E+06 &  2.51 & 0.7E+06 &  24.35 & 2.8E+06 &  63.05 & 4.3E+06 & 0.39 &  4.63 $ \pm $ 0.20 \\
C19          & 0.77 & 3.7E+06 &  1.87 & 4.3E+06 &  16.38 & 1.6E+07 &  32.32 & 1.9E+07 & 0.51 &  3.11 $ \pm $ 0.21 \\
C20 (C19)    & 0.79 & 1.6E+06 &  1.90 & 1.9E+07 &  14.00 & 5.8E+06 &  28.65 & 7.1E+06 & 0.49 &  2.66 $ \pm $ 0.21 \\
A28 (C20)    & 1.30 & 0.3E+06 &  3.42 & 0.3E+06 &  19.26 & 0.8E+06 &  39.05 & 1.0E+06 & 0.49 &  3.66 $ \pm $ 0.21 \\
A29 (C20)    & 1.01 & 0.6E+06 &  2.66 & 0.8E+06 &  14.18 & 1.8E+06 &  29.49 & 2.2E+06 & 0.48 &  2.69 $ \pm $ 0.21 \\
C21          & 1.02 & 1.7E+06 &  1.74 & 1.4E+06 &  14.49 & 4.8E+06 &  32.92 & 6.6E+06 & 0.44 &  2.75 $ \pm $ 0.21 \\
A30 (C21)    & 1.01 & 0.2E+06 &  2.40 & 0.2E+06 &   7.42 & 0.3E+06 &  22.65 & 5.6E+06 & 0.33 &  1.41 $ \pm $ 0.22 \\
C22          & 1.28 & 5.7E+06 &  2.41 & 5.2E+06 &  26.67 & 2.4E+07 &  50.59 & 2.7E+07 & 0.53 &  5.07 $ \pm $ 0.20 \\
A31 (C22)    & 1.53 & 1.0E+06 &  3.05 & 1.0E+06 &  31.41 & 4.4E+06 &  56.30 & 4.7E+06 & 0.56 &  5.98 $ \pm $ 0.20 \\ 
\hline
\end{tabular}
\label{IrasFluxes}
}
\end{center}
\end{table*}   

\section{Acknowledgements}

The authors wish to thank the British 
Council and the General Secretariat of Research and Technology for financial 
support. M. Kontizas would like to thank the University of Athens (ELKE) for 
partial financial support. Finally M. Kontizas, E. Livanou and I. Gonidakis 
are very much indebted to the Ministry of Education for the financial support 
through the ``Pythagoras II'' program.

\end{document}